\documentclass[
  aps,
  prb,
  reprint,
  superscriptaddress,
  longbibliography,
  showkeys,
  citeautoscript,
]{revtex4-1}

\usepackage[intlimits]{amsmath}
\usepackage{amssymb}
\usepackage{amsthm}
\usepackage{booktabs}
\usepackage{braket}
\usepackage{color}
\usepackage{float}
\usepackage{graphicx}
\usepackage{glossaries}
\usepackage[normalem]{ulem}
\usepackage[T1]{fontenc}
            
\usepackage[colorlinks=true,
            linkcolor=black,
            citecolor=black,
            filecolor=black,
            urlcolor=black
            ]{hyperref}
\usepackage[utf8]{inputenc}
\usepackage[version=4]{mhchem}
\usepackage{multirow}
\usepackage{soul}
\usepackage{tabularx}
\usepackage{units}
\usepackage[dvipsnames]{xcolor}

\usepackage{dcolumn}
\usepackage{float}
\usepackage{booktabs}

\setacronymstyle{long-short}
\newacronym{qd}{QD}{quantum dot}
\newacronym{pl}{PL}{Photoluminescence}
\newacronym{cl}{CL}{Cathodoluminescence}
\newacronym{gaas}{GaAs}{Gallium Arsenide}
\newacronym{ingaas}{InGaAs}{Indium Gallium Arsenide}
\newacronym{ebl}{EBL}{Electron beam lithography}

\newacronym{vis}{VIS}{visible}
\newacronym{nir}{NIR}{near-infrared}
\newacronym{uv}{UV}{ultraviolet}
\newacronym{mse}{MSE}{mean squared error}
\newacronym{si}{SI}{Supporting Information}

\newcommand{\dtu}{
    Department of Electrical and Photonics Engineering, Technical University of Denmark,
    2800 Kgs. Lyngby, Denmark
}
\newcommand{\jku}{
    Institute of Semiconductor and Solid State Physics, Johannes Kepler University Linz, Linz, Austria
}

\begin{document}

\title{Deterministic fabrication of GaAs-quantum-dot micropillar single-photon sources}

\author{Abdulmalik A. Madigawa}
\affiliation{\dtu}

\author{{Martin Arentoft Jacobsen}}
\affiliation{\dtu}

\author{Claudia Piccinini}
\affiliation{\dtu}

\author{Paweł Wyborski}
\affiliation{\dtu}

\author{Ailton Garcia Jr.}
\affiliation{\jku}

\author{Saimon F. Covre da Silva}
\affiliation{\jku}

\author{Armando Rastelli}
\affiliation{\jku}

\author{Battulga Munkhbat}
\email[]{bamunk@dtu.dk}
\affiliation{\dtu}

\author{Niels Gregersen}
\email[]{ngre@dtu.dk}
\affiliation{\dtu}

\begin{abstract}

This study investigates the performance of droplet-etched GaAs quantum dots (QDs) integrated into micropillar structures using a deterministic fabrication technique. We demonstrate a unity QD positioning yield across 74 devices and consistent device performance. Under p-shell excitation, the QD decay dynamics within the micropillars exhibit biexponential behavior, accompanied by intensity fluctuations limiting the source efficiency to < 4.5\%. Charge stabilization via low-power above-band LED excitation effectively reduces these fluctuations, doubling the source efficiency to $\sim$ 9\%. 
Moreover, we introduce suppression of radiation modes by introducing cylindrical rings theoretically predicted to boost the collection efficiency by a factor of 4. Experimentally, only a modest improvement is obtained, underscoring the influence of even minor fabrication imperfections for this advanced design.
Our findings demonstrate the reliability of our deterministic fabrication approach in producing high-yield, uniform devices, while offering detailed insights into the influence of charge noise and complex relaxation dynamics on the performance.

\end{abstract}

\maketitle

\section{Introduction}

The development of high-quality single-photon sources (SPSs) with near-unity single-photon purity, efficiency, and indistinguishability is fundamental to advancing photonic quantum information processing applications such as photonic quantum computing \cite{Knill2001AOptics}, secure quantum communication \cite{Gisin2007QuantumCommunication}, quantum networks \cite{Wehner2018QuantumAhead} and quantum sensing \cite{Pirandola2018AdvancesSensing}. Self-assembled quantum dots (QDs) are particularly promising for on-demand single-photon generation, as confined excitons behave as two-level systems capable of  deterministic emission, in contrast to probabilistic sources based on spontaneous parametric down-conversion (SPDC)\cite{Shan2013SingleDots}. Additionally, their compatibility with conventional semiconductor micro- and nanofabrication processes makes them advantageous for integration into photonic devices and scalable quantum photonic processors. Moreover, QDs embedded in microcavities (waveguides) benefit from cavity (waveguide) quantum electrodynamics effects, enhancing both the photon extraction efficiency and the indistinguishability \cite{Heindel2023QuantumTechnology, Lodahl2015InterfacingNanostructures}. Among various monolithic SPS  designs, vertical micropillar cavities currently offer the highest performance thanks to strong Purcell enhancement combined with highly directional emission\cite{Maring2024APlatform, Wang2020MicropillarIndistinguishabilityb, Wang2019, Unsleber2016HighlyEfficiency, Somaschi2016}. 

Nevertheless, the realization of high-performance SPS devices remains challenging. The self-assembled growth process of QDs inherently produces a random spatial and spectral distribution \cite{Bart2022Wafer-scaleDensity}, creating difficulties in positioning QDs precisely within the microcavity structure, which limits collection efficiency and reduces device yield. Deterministic fabrication techniques, including two-color photoluminescence \cite{Sapienza2015, Madigawa2023AssessingDevices} and in-situ optical \cite{Dousse2008} and electron-beam lithography \cite{Gschrey2013}, have been developed to address this issue, enabling the creation of high-performance devices \cite{Liu2019AIndistinguishability, Li2023ScalableCircuit, Somaschi2016}. However, discrepancies between predicted and experimentally achieved Purcell factors and source efficiencies are still significant due to fabrication imperfections. Second, the solid-state nature of the QD and its surrounding matrix introduces dephasing due to spectral diffusion caused by charge fluctuations and phonon-induced dephasing, both of which reduce photon indistinguishability \cite{Senellart2017}. Various approaches to mitigate these effects have been explored, including charge stabilization via an external electric field \cite{Schimpf2021Entanglement-based20K, Zhai2020Low-noisePhotonics}, advanced cavity designs to suppress phonon sidebands \cite{Denning2020PhononSources} and the application of the resonant excitation scheme \cite{Scholl2019ResonanceIndistinguishability, Somaschi2016, Unsleber2016HighlyEfficiency}.

In this study, we investigate the performance of droplet-etched GaAs QDs embedded in micropillar cavities produced using a deterministic fabrication approach for precise QD positioning. This method yields high QD positioning accuracy and consistent performance across fabricated devices. We evaluate the SPS efficiency, decay dynamics, and photon indistinguishability under p-shell pulsed excitation, analyzing discrepancies between theoretical predictions and experimental results. Significant emission intensity fluctuations are observed under p-shell excitation, affecting the source efficiency. To mitigate this, we implement low-power LED illumination for charge stabilization and assess its impact on both efficiency and photon indistinguishability. Additionally, we explore the integration of cylindrical rings around micropillars, a design theoretically predicted to enhance photon extraction by suppressing background modes \cite{Jacobsen2023TowardsModes}. By experimentally assessing the impact of these rings on the collection efficiency and comparing the results with simulations, we identify key factors limiting the device performance.

\begin{figure*}[hbt!]
\includegraphics[width=0.75\textwidth]{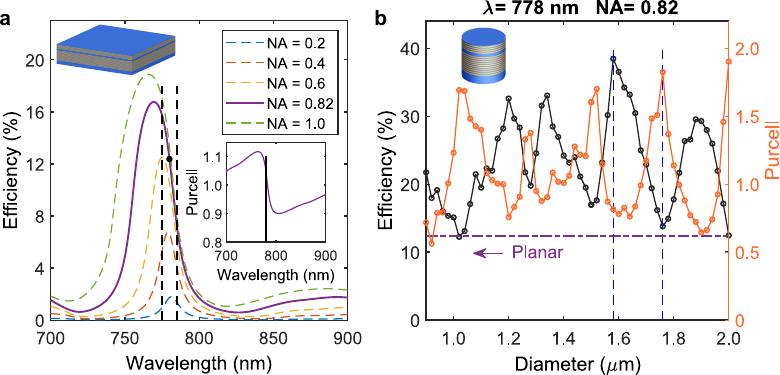}
\caption{Simulation of micropillar structures. (a) Collection efficiency as a function of wavelength of a planar structure for different numerical apertures (NAs). The vertical dashed lines indicate the pre-selected QDs wavelength range (775 to 785 nm). Inset: Purcell factor of the planar structure as a function of wavelength. (b) Collection efficiency (NA = 0.82) and Purcell factor as a function of micropillar diameter at the central wavelength of the QD distribution (778 nm). Purple dashed: Collection efficiency of the planar structure in (a).}
\label{figure 1}
\end{figure*} 

\section{Micropillar Design}

The QD planar sample used in this study consists of a GaAs QD grown via molecular beam epitaxy (MBE) using the droplet etching growth technique\cite{Huo2013Ultra-smallSubstrate}. The QD is embedded at the center of a $\lambda$-cavity, which comprises a $\lambda/2$-thick Al$_{0.33}$Ga$_{0.67}$As layer sandwiched between two $\lambda/4$-thick Al$_{0.2}$Ga$_{0.8}$As layers. The cavity is enclosed by two distributed Bragg Reflectors (DBRs), where the bottom (top) DBR comprises 9.5 (4) pairs of alternating $\lambda/4$-thick Al$_{0.2}$Ga$_{0.8}$As/Al$_{0.95}$Ga$_{0.05}$As layers. Additionally, a 4 nm GaAs cap layer was grown on top of the structure for protection.

We first perform numerical simulations to assess the collection efficiency and Purcell factor as functions of the emission wavelength for a planar structure with the layer sequence described above. Because of experimental uncertainties, the exact layer thicknesses and refractive indices used in these simulations are determined from reflectivity measurements of the planar sample (details in Supporting Note 1). Figure \ref{figure 1}(a) shows the planar structure’s efficiency for various numerical apertures (NAs). The blue shift of the peak efficiency with increasing NA arises from the cavity dispersion effect. With an NA of 0.82, corresponding to our setup’s objective lens, a maximum efficiency of approximately 17\% is achieved. Within the emission range of our QDs (775 nm to 785 nm), indicated by the vertical dashed lines, we expect a collection efficiency of around 12\% and a Purcell factor of $\sim 1$.

To enhance the collection efficiency, we introduce lateral optical confinement by etching micropillars. Figure \ref{figure 1}(b) presents computed efficiency and Purcell factor as functions of the micropillar diameter at an emission wavelength of 778 nm. The efficiency and Purcell factor exhibit opposite trends: peaks in the Purcell factor correspond to dips in efficiency and vice versa. This behavior results from enhanced background emission occurring at the onset of new guided modes \cite{Wang2021SuppressionCavities}. These background modes are not captured by the chosen NA, leading to decreased efficiency. As the Purcell factor remains relatively low ($\sim$ 1), background modes significantly influence the overall Purcell factor. Specifically, these background modes provide additional emission channels for the emitter, causing the visible peaks in Fig.\ \ref{figure 1}(b). The vertical dashed blue lines highlight two diameters $D$ of interest: $D = 1.58$ $\mu$m, with a peak efficiency of approximately 39\%, and $D = 1.76$ $\mu$m, with a peak Purcell factor of $\sim$ 1.8. These diameters are the target sizes for fabrication. Additional simulations for various emission wavelengths and diameters are detailed in Supporting Note 2. \\

\section{Deterministic integration of QDs in micropillar cavities}

\begin{figure*}[hbt!]
\centering\includegraphics[width=0.9\textwidth]{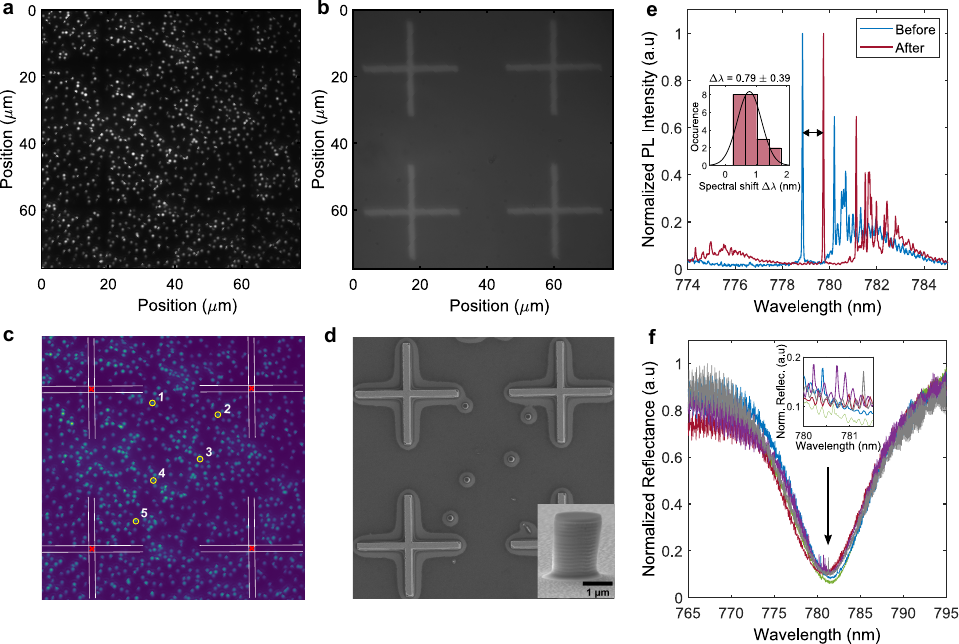}
\caption{Deterministic fabrication of single-QD micropillar structures. (a) Photoluminescence image of the QDs in a planar structure and (b) image of alignment markers taken with a confocal imaging setup. (c) Processed image of the combined QDs and markers showing the extracted location of the alignment markers and pre-selected QDs. (d) Scanning electron microscope image showing the fabricated micropillar structures containing a single QD. (e) Example of photoluminescence spectrum of one of the pre-selected QDs before and after structuring, showing the spectral red-shift of the emission peak. Inset: Histogram of the spectral shift distribution. (f) Reflectance intensity of five different devices, showing the QDs spectrally positioned within the cavity mode. Inset: Zoomed-in plot of the QDs peaks.}
\label{figure 2}
\end{figure*}

To achieve the predicted device performance, the QD should be centered in the micropillar structure. For this purpose, we employ a deterministic fabrication method based on the photoluminescence (PL) imaging technique \cite{Madigawa2023AssessingDevices} to ensure accurate positioning of the QD relative to the micropillar’s center. We begin by fabricating an array of alignment markers on the planar sample through standard lithography and lift-off processes (see Supporting Note 3 for more details). The sample is then cooled to 4 K in a closed-cycle cryostat, where we map the PL spectra of the QDs within the region enclosed by the alignment markers. For further processing, we select only QDs that are spectrally aligned with the central range of the planar cavity mode and spatially isolated from neighboring QDs. To determine the precise coordinates of the pre-selected QDs, we utilized the two-color confocal PL imaging technique \cite{Madigawa2023AssessingDevices}. The merged-images approach is used, in which separate images of the QDs' PL and alignment markers are taken and then combined during image processing (Fig.\ \ref{figure 2}(a) and Fig.\ \ref{figure 2}(b)). This acquisition approach is preferred over the single-image approach as it mitigates alignment errors caused by the high density of QDs near the markers, which can interfere with accurate marker center detection. We use an image analysis program consisting of edge detection and Gaussian blob detection algorithms to localize the alignment markers and the QDs, respectively \cite{Madigawa2023AssessingDevices}, as shown in Fig.\ \ref{figure 2}(c). The extracted QD coordinates, referenced to the alignment markers, are transformed into global sample coordinates and subsequently used to fabricate the micropillar devices via electron-beam lithography and dry etching processes. To address potential size variations during fabrication, which may shift the micropillar dimensions away from their optimal values, we fabricated micropillars with slightly varied diameters within the optimum size range. The etching profile was optimized to minimize the sidewall inclination and corrugation. A scanning electron microscope image of one of the fabricated micropillars is shown in Fig.\ \ref{figure 2}(d), demonstrating straight sidewalls with minimal corrugation.

To verify that the fabricated micropillars contain the pre-selected QDs, we perform PL measurements on all fabricated devices using a low temperature (4 K) microphotoluminescence ($\mu$-PL) setup (see Supporting Information Note 4 for setup details). By comparing the PL spectra recorded before and after fabrication, we confirm that all 74 fabricated devices successfully contain the selected QDs, achieving a spatial positioning yield of unity. Although direct measurements of the alignment offset from the micropillar center are not performed in this study, our previous work using the same method \cite{Madigawa2023AssessingDevices} demonstrates that the QDs are aligned to the center of the structures with <100 nm alignment offset. This level of offset has minimal impact on the efficiency for the micropillar diameters investigated here\cite{Madigawa2023AssessingDevices}. Figure \ref{figure 2}(e) shows the PL spectrum for one and the same QDs before and after micropillar fabrication, measured under above-band pulse excitation at 660 nm. Notably, a redshift of the neutral exciton emission peak (the most pronounced peak in the spectra) is observed, with a mean shift of $(0.79 \pm 0.39)$ nm across devices, as shown in the inset of Fig.\ \ref{figure 2}(e). Such shifts, also reported in other studies \cite{Pregnolato2020DeterministicDots, Kaganskiy2015AdvancedProcessing, Jons2011DependenceStress}, are often attributed to changes in the QD electrostatic environment or fabrication-induced stress. Specifically, charged surface states at the pillar edges may induce lateral electric fields at the QD position. By polarizing the confined excitons, such electric fields would produce a redshift of the QD emission. In addition, etching followed by exposure of the pillars to air causes partial oxidation of the Al-rich layers in the DBRs, which we expect to produce tensile stress in the QD layer and, consequently, a redshift of the emission. 


To assess the spectral alignment of the QDs within the cavity mode, we measure the reflectance spectrum of the micropillars using a white light source. The source is coupled to a single-mode fiber and focused through a microscope objective to a spot size of $\sim$ 2 $\mu$m. Simultaneously, the source excites the QD, enabling us to observe the QD PL within the cavity mode spectrum, as shown in Fig.\ \ref{figure 2}(f). Due to the spectral distribution of the QDs and the shift observed after fabrication, the yield of devices with QDs spectrally aligned to the cavity mode center depends on the design’s flexibility. In this study, the use of a low-Q cavity allows for good QD-cavity spectral alignment without additional tuning. However, for high-Q cavities with narrow cavity modes, spectral tuning methods such as Stark tuning may be needed to precisely align the QD emission with the cavity mode center \cite{Najer2019AMicrocavity, Somaschi2016}.

\begin{figure*}[hbt!]
\centering\includegraphics[width=0.7\textwidth]{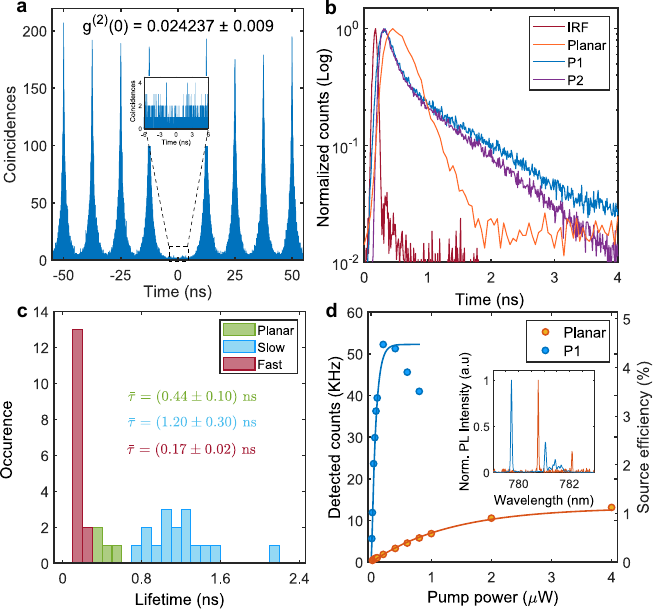}
\caption{Single-photon emission properties of QDs in a micropillar structure. (a) Second-order correlation of a device's emission under p-shell pulse excitation. (b) Time-resolved photoluminescence measurements of representative QDs in planar and micropillar devices (P1 ($D = 1.58$ $\mu$m) and P2 ($D = 1.76$ $\mu$m)) under p-shell excitation. IRF is the instrument response function of the SNSPD measured with a pulsed laser at the wavelength of the QD emission. The measurements were taken with a 20 ps time bin. (c) Histogram of decay times of various QDs in planar and micropillar devices. The quoted decay times are obtained from a mono and bi-exponential function fit for the planar and micropillar samples, respectively. (d) Detected count rate as a function of pump power for a planar and the P1 device. The right axis shows the source efficiency at the first lens obtained after accounting for the setup detection efficiency. Solid curves represent fits to the saturation curve \(C_{0} (1 - \text{exp}(-P / P_{0}))\). Inset: the photoluminescence spectrum of the measured devices.}
\label{figure 3}
\end{figure*}

\section{Optical characterization}

For detailed characterization, we selected two micropillar devices: P1 ($D = 1.58$ $\mu$m), corresponding to the optimal efficiency prediction, and P2 ($D = 1.76$ $\mu$m), which exhibits the peak Purcell factor but lower efficiency, as indicated in Fig.\ \ref{figure 1}(b). The samples are excited using a picosecond pulsed Ti:Sapphire laser (80 MHz repetition rate, $\sim$ 4 ps pulse width), tuned to one of the QD’s excited states, referred to as the p-shell resonance. The emitted photons are spectrally filtered using a grating spectrometer slit (equivalent to $\sim$ 0.2 nm spectral window) and then coupled to a superconducting nanowire single-photon detector (SNSPD). We measured and analyzed key device performance metrics, including emitter decay dynamics, single-photon purity, brightness, and photon indistinguishability.

\subsection{Purity and Decay dynamics}

To assess the single-photon emission characteristics of the fabricated micropillars, we measure the second-order autocorrelation \(g^{(2)}(\tau)\) using a standard Hanbury Brown and Twiss setup. Figure \ref{figure 3}(a) shows the coincidence counts histogram for a representative micropillar (P1) under p-shell pulsed excitation (detuning $\sim$ 9 meV) at \(P= P_{sat}\). The absence of coincidence counts at zero delay confirms single-photon emission, with a \(g^{(2)}(0)\) value of $0.024 \pm 0.009 $. This value is obtained by calculating the ratio of the central peak area to the average area of the surrounding peaks. The uncertainty reflects the statistical variation of the surrounding peaks, representing one standard deviation under the assumption of a Poisson distribution of total counts. No blinking or carrier recapturing signature is observed in the \(g^{(2)}(\tau)\) for the measured timescale (up to 500 ns time delay). This is in contrast to measurements performed using above-band pulse excitation showing bunching around zero delay and at a longer time scale ($\sim$ 100 ns) (see Fig.\ S6, Supporting Information).

Next, we investigate the spontaneous emission decay dynamics of the fabricated devices using time-resolved PL measurements. Figure \ref{figure 3}(b) compares the decay curves for the neutral excitons confined in QDs in the planar and micropillar structures, showing a clear transition from mono-exponential decay in the planar sample to bi-exponential decay in the micropillar devices. A histogram of the decay times across various fabricated micropillars is presented in Fig.\ \ref{figure 3}(c). The planar sample exhibits a monoexponential decay with an average decay time of \(\tau= (0.44 \pm 0.1)\) ns. In contrast, the micropillar devices exhibit two distinct decay components: a fast decay time of \(\tau= (0.17 \pm 0.02)\) ns, approximately half the planar sample’s decay time, and a slow decay time of \(\tau= (1.20 \pm 0.3)\) ns, roughly three times longer than that of the planar sample. The fast decay dynamics are consistent across all deterministically fabricated devices, as evidenced by the narrow lifetime distribution. Contrarily, measurements on randomly fabricated devices with unaligned QDs exhibit significant variation (see Fig.\ S7, Supporting Information), highlighting the success and reproducibility of our deterministic fabrication approach.


In a related study on similar QDs in a planar structure, significantly slower decay times (as long as 1.8~ns) are observed under p-shell or above-band excitation compared to resonant or longitudinal-acoustic (LA) phonon-mediated excitation (typically about 200~ps) \cite{Reindl2019HighlyDots}. This slow relaxation is typically observed in large QDs with densely spaced excited states and has first been attributed to restrictive orbital/spin selection rules governing the relaxation process \cite{Reindl2019HighlyDots, Jahn2015AnLinewidth} and, more recently, to the reduced phonon spectral density at the typical energy spacing of $\sim3$~meV between confined hole states~\cite{Lehner2023}. In other studies, the biexponential nature of a QD emission is attributed to the coupling of dark states with bright excitonic states through phonon-mediated spin-flip processes, resulting in fast and slow decay components associated with the bright and dark exciton states, respectively \cite{Madsen2016RoleCavities, Tighineanu2016Single-PhotonDot}. In this study, the origin of the bi-exponential decay observed in the micropillar structures is likely due to one of the above mechanisms.

\subsection{Source efficiency}

\begin{figure*}[hbt!]
\centering\includegraphics[width=0.9\textwidth]{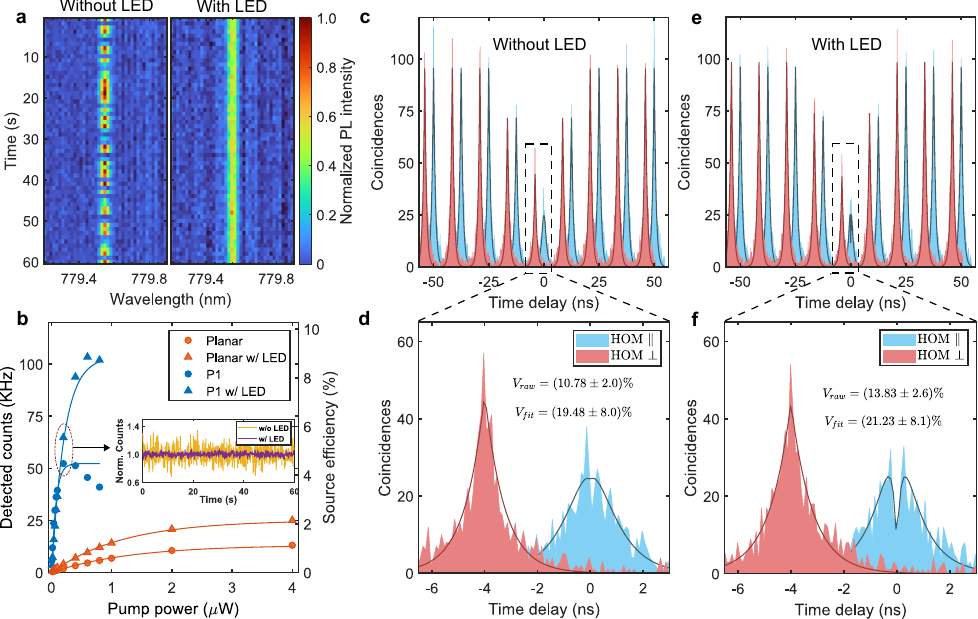}
\caption{Charge stabilization of QD under p-shell excitation with additional above-band LED excitation. (a) Photoluminescence map as a function of time for device P1, showing the intensity fluctuations of the emitter under p-shell excitation (left) and the stabilization of the emitter under additional low-power above-band LED excitation (right). For each frame (1 s), the PL spectrum was acquired for 0.25 s at pump power $P= 10$ nW. (b) Detected count rate and source efficiency as a function of pump power for the planar and micropillar device P1. Solid curves represent fits to the saturation curve \(C_{0} (1 -\exp(-P / P_{0}))\). Inset: Detected count rate as a function of time (integration time = 100 ms) for the measurement with and without LED, normalized to the average count of each measurement. (c, d) and (e, f) Hong-Ou-Mandel two-photon interference experiment  coincidence histograms taken at \(P=2P_{0}\) for measurements with and without LED, respectively (bin width = 100 ps). The visibility is obtained by integrating the data over the full 12.5 ns window. Solid lines correspond to fit using a double-sided mono-exponential function. For clarity, a temporal offset of 4 ns was added between the parallel (HOM$_{\parallel}$) and orthogonal (HOM$_{\perp}$) polarization measurements. The uncertainty on the raw visibilities was calculated from the assumption of a Poissonian distribution of the total counts of each peak.}
\label{figure 4}
\end{figure*}

We characterize the source efficiency by measuring the total detected counts per second registered on a single-photon detector, and Fig.\ \ref{figure 3}(d) shows the average count rate as a function of the excitation laser power for QDs in a micropillar and in the planar sample. We estimate the source efficiency as the fraction of the excitation pulses received at the first lens, taking into account the overall transmission of the optical setup (\(\eta_\text{setup} = 1.43 \%\)) (see Supporting Note 5 and Table S1). At its maximum, the P1 micropillar ($D = 1.58$ $\mu$m) measurement features approximately four times the collection efficiency of the planar sample, in good agreement with simulation predictions. However, quantitatively, the source efficiency of $\sim$ 4.5\% falls short of theoretical expectation. Furthermore, we observe significant fluctuations in the emission intensity for different devices, suggesting an unstable QD charge environment. Figure \ref{figure 4}(a) (left) depicts these fluctuations in the emission intensity, likely caused by impurities or defects in the QD that introduce residual charges, leading to fluctuations between the neutral and charged states of the QD \cite{Reindl2019HighlyDots}. These fluctuations are partly stabilized by introducing an additional low-power above-band excitation with a light-emitting diode (LED) emitting at 470 nm, as shown in Fig.\ \ref{figure 4}(a) (right). As reported in previous studies, this approach stabilizes the QD charge environment by filling up possible charge traps or defect states near the QD, resulting in stabilization of the emission intensity
\cite{Nguyen2013PhotoneutralizationSpectroscopy, Gazzano2013, Houel2012ProbingDot, Huber2016EffectsDots}. The LED power was set low enough to minimize its contribution to the overall PL counts while effectively reducing the fluctuations (see Fig.\ S8(a), Supporting Information). Figure \ref{figure 4}(b) shows the count rate as a function of excitation power for QDs in the P1 micropillar and in the planar structure with/without the additional LED excitation. As a consequence of stabilizing the charge environment, the LED excitation increases the maximum count rate, highlighting the impact of charge noise on the source efficiency. The decay dynamics and \(g^{(2)}(0)\) remain unchanged in the micropillar with LED excitation, and again no evidence of recapturing is seen (Fig.\ S8(b), Supporting Information). The inset in Fig.\ \ref{figure 4}(b) displays the normalized count rate over time, demonstrating a reduction in the detected count fluctuations by nearly 50\% with the LED excitation.

\begin{figure*}[hbt!]
\centering\includegraphics[width=0.7\textwidth]{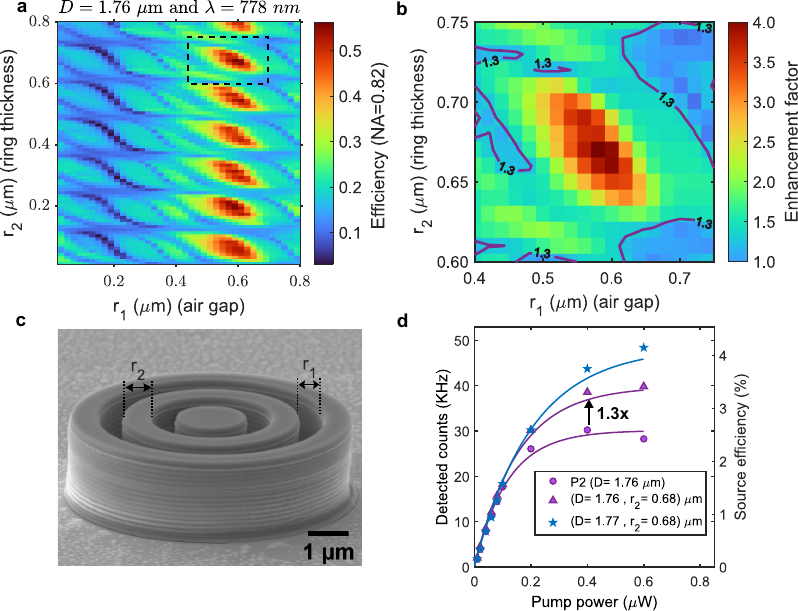}
\caption{Investigation of the effect of adding 2 rings around a micropillar. (a) Simulated collection efficiency as a function of ring thicknesses and air gap. (b) Collection efficiency enhancement produced by the rings over a bare micropillar for the region defined by the black dashed box in (a). The contour lines indicate the experimentally measured enhancement factor. (c) Scanning electron microscope image of the fabricated rings around the micropillar, targeting the highlighted region in (a). (d) Detected count rate and source efficiency as a function of pump power for micropillars with and without rings under p-shell excitation, with additional LED excitation for charge stabilization. The rings were fabricated with a r$_1$ = 0.6 $\mu$m fixed airgap. }
\label{figure 5}
\end{figure*}

\subsection{Photon indistinguishability}

To assess the indistinguishability of the source and examine the impact of charge stabilization via LED illumination on photon coherence, we performed a Hong-Ou-Mandel (HOM) two-photon interference (TPI) experiment on device P1\cite{Hong1987MeasurementInterference, Santori2002IndistinguishableDevice}. The micropillar is excited with an 80 MHz repetition (12.5 ns period)  pulsed laser tuned to the p-shell resonance of the QD. The emitted single photons are spectrally filtered and coupled to an unbalanced Mach-Zehnder interferometer (MZI: see Supporting Note 6 and Fig.\ S4 for setup details). A 12.5 ns fiber delay loop and a polarization controller are introduced to one arm of the interferometer to temporally align two consecutively emitted photons so that they interfere at a 50/50 beam splitter (BS). Perfectly indistinguishable photons would always exit the BS through the same output, leading to zero coincidence counts at zero delay. A half-wave plate in one arm is used to switch the interfering photons between parallel ($\parallel$) and orthogonal ($\perp$) polarization configurations, making the photons either nominally indistinguishable or distinguishable, respectively. Coincidence histograms for parallel (HOM$_{\parallel}$) and orthogonal (HOM$_{\perp}$) polarizations from the TPI experiment for the P1 micropillar with and without LED excitation are presented in Figs.\ \ref{figure 4}(c) and \ref{figure 4}(e), respectively. The reduced central peak area of the HOM$_{\parallel}$ measurement compared to the HOM$_{\perp}$ measurement is a clear signature of quantum interference, and the visibility contrast between the two measurements quantifies the degree of photon indistinguishability. The raw TPI visibility is calculated from the ratio of the central peak areas in the co- and cross-polarized cases via \({V}_\text{raw} = 1-(A_{\parallel}/ A_{\perp})\), where the areas are normalized to the average of surrounding peaks, excluding the first (see Supporting Information Note 6 for more details). The extracted raw visibility is $(10.8 \pm 2.0) \% $ without the additional LED excitation. This value increases to $(13.8 \pm 2.6) \%$ with LED excitation with a pronounced dip at zero delay. A maximum visibility value of $(21 \pm 8) \% $ is extracted by fitting the data with a double-sided mono-exponential decay function (see Supporting Information Note 7 for the fitting details). 
For p-shell excitation, the complex and slow relaxation dynamics involving the excited states and possibly dark states discussed above most likely limit the indistinguishability of two subsequently emitted photons by introducing a significant time jitter in the creation time of an exciton relative to the laser pulse~\cite{Reindl2019HighlyDots}. 
The observed increase in the dip depth at zero delay with the LED as compared to without LED (Figs.\ \ref{figure 4}(d) and \ref{figure 4}(f)) suggests an improvement in the mutual coherence of the photons. This can be attributed to the reduction of spectral diffusion or timing jitter due to a more stable charge environment. However, despite stabilization, the LED excitation does not fully mitigate charge noise or other dephasing mechanisms \cite{Gerhardt2018IntrinsicSource}, which -- together with the slow relaxation dynamics \cite{Reindl2019HighlyDots} --  collectively contribute to the observed low HOM visibility. For a perfect dephasing- and time-jitter-free system, the coherence time \(\tau _{c}= 2\tau\), resulting in a Fourier-transform-limited emission \cite{Santori2002IndistinguishableDevice}. From the central dip width in Fig.\ \ref{figure 4}(f), we extract a coherence time of $\tau_{c} = (219 \pm 17)$ ps, which is somewhat shorter than the expected coherence time of the radiative decay of the bright exciton (about 400~ps) for these QDs. High HOM visibility can be expected by eliminating the slow relaxation dynamics via alternative excitation schemes, such as longitudinal acoustic (LA) phonon-assisted or resonant excitation \cite{Reindl2019HighlyDots, Scholl2019ResonanceIndistinguishability}. 


\section{Micropillar with rings}

As demonstrated by Jacobsen et al. \cite{Jacobsen2023TowardsModes}, cylindrical rings surrounding a micropillar structure can increase collection efficiency by suppressing background modes. In this study, we experimentally investigate the impact of incorporating such rings on the collection efficiency of micropillars. For this purpose, we selected the 1.76 $\mu$m diameter micropillar, corresponding to a local minimum in the collection efficiency curve shown in Fig.\ \ref{figure 1}(b), aiming to enhance the efficiency by adding rings. We carried out numerical simulations of the micropillar surrounded by two cylindrical rings with identical ring thickness (r$_{2}$) and air gap (r$_{1}$). The computed collection efficiency as a function of r$_{1}$ and r$_{2}$  is presented in Fig.\ \ref{figure 5}(a). A quasi-periodic pattern appears as a function of r$_{2}$, showing periodic enhancements in the collection efficiency. This behavior suggests that the rings effectively suppress background modes, thereby channeling the emission into the fundamental mode \cite{Jacobsen2023TowardsModes}. The improvement is quantified using an enhancement factor, defined as the ratio of the collection efficiency of the micropillar with rings to that of the micropillar without rings. Figure \ref{figure 5}(b) presents the enhancement factor for the parameter space indicated by the dashed box in Fig.\ \ref{figure 5}(a).  The results reveal a maximum collection efficiency enhancement by a factor of 4, demonstrating the rings' ability to suppress background modes and improve the collection efficiency significantly.

The devices are then fabricated targeting optimized ring parameters (corresponding to the peak in Fig.\ \ref{figure 5}(b)) using the deterministic fabrication procedure described earlier. To account for fabrication imperfection, multiple devices were fabricated with slight variations in micropillar diameter and ring thickness while maintaining a fixed air gap of r$_1$ = 0.6 $\mu$m. Figure \ref{figure 5}(c) shows a scanning electron microscope (SEM) image of a fabricated micropillar with two rings. The detected count rates for micropillars with and without rings were compared, as shown in Fig.\ \ref{figure 5}(d). The micropillar with rings exhibited an $\sim$ 1.3$\times$ enhancement in detected count rates relative to the micropillar without rings. However, this improvement is significantly lower than the $\sim$ 4$\times$ enhancement predicted by the simulations. This discrepancy is likely due to deviations between the fabricated device dimensions and the intended design parameters, causing the device to operate in a region of lower enhancement, as indicated in the contour plot of Fig.\ \ref{figure 5}(b). Notably, a higher count rate was observed from a device with a slightly larger pillar diameter, as shown in Fig.\ \ref{figure 5}(d), highlighting the sensitivity of the design to size variations. Moreover, the QD lifetimes measured for devices with and without rings were similar and within the distribution of lifetimes for the devices shown in Fig.\ \ref{figure 3}(c).

\section{Discussion}

Using a deterministic fabrication approach, we have successfully integrated QDs into the center of micropillar structures with unity yield and positioning accuracy of < 100 nm. Assuming 100\% population inversion of the two-level system, we expect a source efficiency of 40\% from the optimum micropillar ($D = 1.58$ $\mu$m) as predicted in Fig.\ \ref{figure 1}. However, initial measurements under p-shell excitation show a maximum source efficiency of $\sim$ 4.5 \%, an order of magnitude lower than predicted. Significant fluctuations in the count rate are observed, which is a limiting factor for the source efficiency. This behavior, also present in the planar structure, suggests that the fluctuations are not due to micropillar fabrication imperfections. Instead, they likely arise from an unstable charge environment within the QD, which disrupts the emission process and reduces source efficiency. For an ideal two-level system, the emission rate should plateau beyond saturation. However, the detected count rate as a function of incident laser power, shown in Fig.\ \ref{figure 3}(d), exhibits a sharp drop at higher excitation powers. This suggests that non-radiative processes such as Auger recombination, phonon interactions, or charge noise are dominant at high excitation powers, preventing the emitter from reaching full population inversion and thus reducing the photon emission rate \cite{Kurzmann2016AugerTransition, Heitz1997EnergyDots}. We stabilized the charge environment using above-band LED excitation to address the charge fluctuations. This approach improved the emission rate, achieving $\sim$ 9\% source efficiency, which is double the value without stabilization. While this method effectively reduces charge noise in the surroundings of the QD and in the QD itself and enhances efficiency, it does not fully eliminate all sources of inefficiency. Non-radiative processes, particularly phonon interactions, likely remain dominant, continuing to limit the maximum achievable efficiency.


Time-resolved photoluminescence measurements revealed a transition from a single-exponential decay in planar devices to a bi-exponential decay in structured micropillars under both above-band and p-shell excitation. This shift in decay dynamics implies the presence of additional relaxation and recombination pathways in the micropillar structures, likely related to the enhanced light-matter interaction and changes in the excited states used for excitation. At present we can only speculate on the origin of the observed bi-exponential decay observed in micropillars. We see that both the rise time and the initial decay are faster than in the planar structure. In addition, the initial decay time (about 170~ps) is slightly shorter than the typically observed radiative decay times of the neutral exciton under resonant excitation (about 200~ps) and is followed by a slow decay reminiscent of dark exciton population. We recall here that our "p-shell" excitation is populating one or more of the states that consist, in a single-particle picture, of a ground-state electron and an excited-state hole. On one hand, the moderate Purcell enhancement could accelerate the radiative decay of such excited states, leading to their depopulation before phonon-mediated relaxation to the ground state. On the other hand, possible anisotropic electric and strain fields induced by the processing could sensitively affect the nature of the excited states (in particular, the heavy-hole--light-hole mixing) and their probability and rate of relaxation to the bright or dark ground states. The biexponential decay dynamics point out to an accelerated relaxation/decay of the excited states to the ground state and an enhanced population of the dark states. Both effects could stem from enhanced valence-band mixing of the excited states.



The Hong-Ou-Mandel experiment reveals two-photon interference visibility of $\sim$ 14\% after stabilization of the charge environment using an above-band LED. This low HOM visibility suggests the presence of several mechanisms severely impacting the photon indistinguishability. A major factor contributing to this low visibility is the slow phonon-assisted relaxation of the system, which introduces substantial time jitter in the photon emission. By reducing the integration time window, we were able to extract a HOM visibility as high as 50\% (see Supporting Information Note 11 and Fig.\ S9). This improvement highlights the critical impact of time jitter on photon indistinguishability. One viable route to improving indistinguishability is employing phonon-assisted or resonance excitation schemes, as reported by\cite{Reindl2019HighlyDots}. These approaches eliminate the slow relaxation process, enabling more coherent photon emission and significantly enhancing photon indistinguishability. Moreover, the charge stabilization using the LED does not completely eliminate the charge noise. Here, employing electrical contacts for charge stabilization via an external electric field would provide more robust charge noise suppression, further improving photon indistinguishability.

We incorporated cylindrical rings around the micropillars to improve collection efficiency using  suppression of emission into background radiation modes \cite{Jacobsen2023TowardsModes}. Simulations indicate that adding rings to the low-efficiency micropillar ($D = 1.76$ $\mu$m) should enhance the collection efficiency of the micropillar by a factor of four. However, experimental results reveal only a 1.3$\times$ enhancement in collection efficiency. This discrepancy is likely due to deviations in the fabricated pillar and ring dimensions from the optimal design parameters, as well as fabrication-related losses not accounted for in the simulations. As illustrated in Fig.\ \ref{figure 5}(d), even a 10 nm increase in the micropillar diameter results in $\sim$ 20\% improvement in collection efficiency, underscoring the sensitivity of the design to even modest diameter variations. Further experimental demonstration of these suppression effects will thus require improved control of the fabrication imperfections.

\section{Conclusion}

Using a deterministic fabrication approach, we investigated the performance of single-photon sources based on droplet-etched GaAs quantum dots (QDs) embedded in micropillars. Photoluminescence measurements confirmed that all pre-selected QDs are successfully incorporated resulting in a 100\% yield. While the position accuracy is expected to be below 100 nm, the source efficiency fell significantly short of theoretical predictions. This may be due to slow QD relaxation dynamics resulting from the employed p-shell excitation method, which adversely impacted both efficiency and photon indistinguishability. Charge stabilization via low-power above-band LED excitation improves indistinguishability and doubles the source efficiency, however challenges related to the slow relaxation dynamics persist. We investigated the enhancement of collection efficiency from suppression of the background emission by introducing two cylindrical rings around a micropillar. While simulations predict a $\sim$ 4$\times$ enhancement, we experimentally measured only a $\sim$ 1.3$\times$ enhancement. These results highlight the sensitivity of the ring design to fabrication imperfections and underscore the need for more precise, high-quality fabrication to demonstrate the full predicted enhancement. Overall, our findings highlight the success of our deterministic fabrication in producing uniform, high-yield devices while providing valuable insights into the challenges posed by complex relaxation dynamics and charge noise to achieve optimal device performance.



\section*{Funding}

The authors also acknowledge the European Research Council (ERC-CoG "UNITY", grant no. 865230 and ERC-StG "TuneTMD", grant no. 101076437), the Carlsberg Foundation (grant no. CF21-0496) and the Villum Foundation (grant no. VIL53033).  The authors acknowledge the cleanroom facilities at DTU Nanolab – National Centre for Nano Fabrication and Characterization. Research was also supported by European Union’s Horizon 2020 research and innovation program under Grant Agreement No. 871130 (Ascent+), the EU HE EIC Pathfinder challenges action under grant agreement No. 101115575, the QuantERA II program via the projects QD-E-QKD and MEEDGARD (FFG Grants No. 891366 and 906046) the Austrian Science Fund (FWF) via the Research Group FG5, and the cluster of excellence quantA [10.55776/COE1] as well as the Linz Institute of Technology Secure and Correct Systems Lab, supported by the State of Upper Austria.

\section*{Notes}
The authors declare no competing financial interest.

\section*{Author contributions}

A.G. and S.F.C.D.S grew the QD samples. A.M. fabricated the samples. A.M. performed all the optical characterization of the fabricated samples and data analysis, with support from C.P., P.W., and B.M. M.J. performed optical simulations of the devices. N.G., A.R., and B.M. conceived the idea and coordinated the project. All authors participated in the scientific discussions and manuscript preparation.

\section{References}
\bibliography{citations}

\end{document}